\documentclass[journal]{IEEEtran}
\usepackage{graphicx}
\usepackage{epsfig}
\usepackage{latexsym}
\usepackage{amsfonts}
\usepackage{here}
\usepackage{rawfonts}
\usepackage[latin1]{inputenc}
\usepackage[T1]{fontenc}
\usepackage{calc}
\usepackage{url}
\usepackage{enumerate}
\usepackage{color}
\usepackage[tbtags]{amsmath}
\usepackage{amssymb}
\usepackage{upref}
\usepackage{epic,eepic}
\usepackage{times}
\usepackage{dsfont}
\usepackage{comment}
\usepackage{cite}

\usepackage{graphicx}
\usepackage[font={small}]{caption}
\usepackage[font={small}]{subcaption}

\usepackage{stfloats}
\usepackage{mathtools}
\usepackage{amsmath}

\newtheorem{theorem}{\bf{Theorem}}

\newtheorem{corollary}{\bf{Corollary}}
\newtheorem{lemma}{\bf{Lemma}}

\ifCLASSINFOpdf
\else
\fi
\begin{document}

\title{Fundamentals of Clustered Molecular Nanonetworks}

\author{{Seyed Mohammad Azimi-Abarghouyi}, Harpreet S. Dhillon,~\IEEEmembership{Fellow,~IEEE}, and Leandros Tassiulas,~\IEEEmembership{Fellow,~IEEE}

\thanks{S. M. Azimi-Abarghouyi is with the School of Electrical Engineering and Computer Science, KTH Royal Institute of Technology, Stockholm, Sweden (Email: seyaa@kth.se). H. S. Dhillon is with Wireless@VT, Bradley Department of Electrical and Computer Engineering, Virginia Tech, Blacksburg, VA USA (Email: hdhillon@vt.edu). L. Tassiulas is with the School of Engineering \& Applied Science and the Yale Institute for Network Science, Yale University, New Haven, CT USA (Email: leandros.tassiulas@yale.edu). A part of this work is accepted for presentation at IEEE ICC 2023 \cite{myconf}}.
}

\maketitle

\begin{abstract}
We present a comprehensive approach to the modeling, performance analysis, and design of clustered molecular nanonetworks in which nano-machines of different clusters release an appropriate number of molecules to transmit their sensed information to their respective fusion centers. The fusion centers decode this information by counting the number of molecules received in the given time slot. Owing to the propagation properties of the biological media, this setup suffers from both inter- and intra-cluster interference that needs to be carefully modeled. To facilitate rigorous analysis, we first develop a novel spatial model for this setup by modeling nano-machines as a Poisson cluster process with the fusion centers forming its parent point process. For this setup, we first derive a new set of distance distributions in the three-dimensional space, resulting in a remarkably simple result for the special case of the Thomas cluster process. Using this, total interference from previous symbols and different clusters is characterized and its expected value and Laplace transform are obtained. The error probability of a simple detector suitable for biological applications is analyzed, and approximate and upper-bound results are provided. The impact of different parameters on the performance is also investigated.

\end{abstract}


\section{Introduction}
Molecular communications (MC) has attracted significant research interest due to its numerous applications in biological and communication engineering such as nanoscale sensing, drug delivery, and body area networks \cite{akyildiz}. In MC, biological nano-machines (NMs) release molecules into the environment to realize nano-scale information transfer. Information can be encoded in the concentration or type of the molecules or at the time instants at which they are released. A detector decodes the sent information of an NM based
on the number of captured molecules. Some examples of modulation and inter-symbol interference (ISI) mitigation techniques can be found in \cite{arjmandi_main, reza_main, reza_type, arjmandi_isi}. These works focus on a single pair of transmitting and receiving NMs. On the other hand, nano-networking and Internet of Bio-Nano Things (IoBNT) paradigms require information collection from multiple arbitrarily-located NMs sensing potentially different phenomena \cite{iot}. Hence, interaction and interference of released molecules of multiple NMs over time play a crucial role in the performance characterization and design of bio-nanonetworks. 

Several prior works consider single source interference \cite{evan_fixed} and fixed-location configurations with a limited number of interfering NMs \cite{noel_interference, xia_fixed}. Due to the randomness and irregularity in the locations of NMs, stochastic geometry approaches based on the Poisson point process (PPP) assumption have also been considered for the modeling and analysis of nanonetworks \cite{pier_sg_mc, akan_sg_mc, elkas_sg_mc, dissan_sg_mc, gupta_sg_mc}. In \cite{zabini_sg_mc}, the NMs are modeled as a general point process and the effect of repulsion and attraction is investigated
on the performance using second-order properties of the point process. However, explicit expressions
are only derived for the case when the NMs are distributed as a PPP. These early stochastic geometry-based works lack in two important aspects that inspired this paper. First, these works did not explore the dependence of detectors on interference by considering an interference-dependent decision rule even though interference causes significant increase in the molecule count. Second, despite the usefulness of PPP for modeling the uniform deployments of NMs, it cannot model deployments where fusion centers (FCs) are located at places with high NM density to sense and process their information. The latter is particularly important for capturing the fact that many biological phenomena, such as abnormalities, are localized and multiple cooperative NMs, e.g., each monitoring a different feature of a target, might be required for fast, accurate, and diverse sensing \cite{abnormality}. For example, there may be different tumor cells or other disease sites as targets. Also, such cells should be monitored for
	smart drug delivery. In \cite{roger}, each NM decides individually on the presence of
	the abnormality based on its observation and informs it to the FC for a final decision. A survey on abnormality detection and its bio-applications is available in \cite{survey}. In such deployments, it is important to take into account non-uniformity as well as the correlation that exists between the locations of the NMs and FCs into clusters.

This paper develops a comprehensive approach to modeling, analysis, and design of bio-nanonetworks consisting of multiple NMs clustered around their respective FCs in a three-dimensional (3D) biological media\footnote{Due to the scales of media, NMs, and FCs, 2D space cannot provide enough accuracy\cite{pier_sg_mc, akan_sg_mc, elkas_sg_mc, dissan_sg_mc, gupta_sg_mc, zabini_sg_mc}.}. Our key contributions 
are highlighted next.

\textit{Clustered Transmission:} In each time slot, we consider that only one NM of a cluster transmits its sensed information to its respective FC. This not only results in inter-cluster interference from the simultaneous transmissions in the current time slot but also both inter- and intra-cluster interference from the transmissions corresponding to the previous time slots. 

\textit{Realistic 3D Modeling and Distance Distributions:} A rigorous analysis of this setup required a spatial model that is rich enough to capture its salient features while being tractable enough to enable mathematical analysis of the aforementioned interference and its effect on the system performance. Inspired by this, we develop a novel approach based on Poisson cluster processes (PCPs) \cite[Sec. 3.4]{haenggi_book} as a {\em canonical} choice within the field of stochastic geometry to facilitate tractable modeling and analysis of 3D clustered bio-nanonetworks. Models based on PCPs have recently been applied to study wireless networks over two-dimensional (2D) regions \cite{azimi_cluster1, azimi_cluster2, azimi_cluster3, afshang_cluster1, afshang_cluster2, afshang_cluster4}. In these works, the network usually follows a Thomas cluster process (TCP) \cite[Definition 3.5]{haenggi_book} or Mat{\'e}rn cluster process (MCP) \cite[Definition 3.6]{haenggi_book}. Further, the formalism for establishing distance distributions in PCPs is well-known \cite{math_distance} and has been applied extensively in the recent years to derive key distance distributions for both the TCP and MCP \cite{afshang_distance, afshang_distance2, gupta, azimi_cluster1}.
	
	Based on the proposed model, we characterize relevant distance distributions of 3D PCPs that are inspired by this setup and provide closed-form results for 3D TCPs. In particular, we identify a specific structure for the 3D TCP that provides a remarkably simple distribution function, which was not possible in the  2D TCP as there is an integral in its function.

\textit{Interference-Aware Detector Design:} We design a low-complexity detector for FCs that can efficiently adjust to the interference. For this purpose, we characterize both ISI and interference from interfering NMs in different clusters, and derive the expected value of intra- and inter-cluster interferences.

\textit{Performance Analysis:} We analyze the performance of the detector in terms of the error probability. As a key step in the analysis, we obtain the Laplace transform of intra- and inter-cluster interferences. In addition to exact results, we provide tractable approximations. 

\textit{Design Insights:} We investigate the impact of different parameters of the system model on the the error probability and the expected value of interference. As expected, our analysis reveals that a higher intensity of cluster centers or distance of a reference NM to the center of its cluster has a degrading effect on the performance. Also, increasing the difference of released molecules for different symbols decreases the error probability. Furthermore, there is an optimal time slot duration in terms of the error probability.

\section{System Model}
In this section, we provide a mathematical model of the bio-nanonetwork, including the spatial distribution of the NMs and FCs, the molecular propagation model, and the transmission scheme.
\subsection{Spatial Model}
We consider a 3D clustered bio-nanonetwork as shown in Fig. 1, where the locations of NMs are modeled as a PCP in $\mathbb{R}^3$. A 3D PCP $\Phi$ can be formally defined as a union of offspring points in $\mathbb{R}^3$ that are located around parent points (i.e., cluster centers). The parent point process is a 3D PPP $\Phi_\text{p}$ with intensity $\lambda_\text{p}$, and the offspring point processes (one per parent) are conditionally independent. The set of offspring points of $\mathbf{x} \in \Phi_\text{p}$ is denoted by ${\cal N}^{{\mathbf{x}}}$, such that $\Phi = \cup_{\mathbf{x}\in \Phi_{\rm p}}{\cal N}^{{\mathbf{x}}}$, and the probability density function (PDF) of each element being at a location $\mathbf{y}+\mathbf{x} \in \mathbb{R}^3$ is $f_{\mathbf{Y}}(\mathbf{y})$. After characterizing theoretical results in terms of general distribution $f_{\mathbf{Y}}(\mathbf{y})$,
we specialize the results to the 3D TCP where the points are distributed
around cluster centers according to an independent Gaussian distribution
\begin{align}
f_{\mathbf{Y}}(\mathbf{y}) = \frac{1}{(2\pi)^\frac{3}{2} \sigma^3}\exp\left(-\frac{\|\mathbf{y}\|^2}{2 \sigma^2}\right),
\end{align}
where $\sigma^2$ is the variance of the distribution. 

In the center of each cluster, there is an FC that detects, gathers and processes the transmitted information of the NMs of the same cluster. In our analysis, we will model each FC as a ball of non-zero radius $r_0$ that will place some restrictions on the placement of FCs and NMs. For instance, NMs cannot lie inside the FCs. While the exact analysis of this modified setting is complicated and will lead to significant loss in tractability, we will include the non-zero radius $r_0$ in many components of our analysis, thus capturing these additional restrictions while maintaining tractability.

It is worth noting that the proposed PCP-based setup reduces to a single cluster setup if the parent point process intensity goes to zero. In fact, the analysis of this special case has not yet appeared in the literature, which further reinforces the generality of this work. For brevity, we will not write separate corollaries for the single-cluster case with the understanding that they can be easily deduced from the main results by substituting $\lambda_\text{p} = 0$ (since our analysis will assume the typical FC at the origin, which will be retained even under this limit). 
\begin{figure}[tb!]
\centering

\includegraphics[width =3.4in]{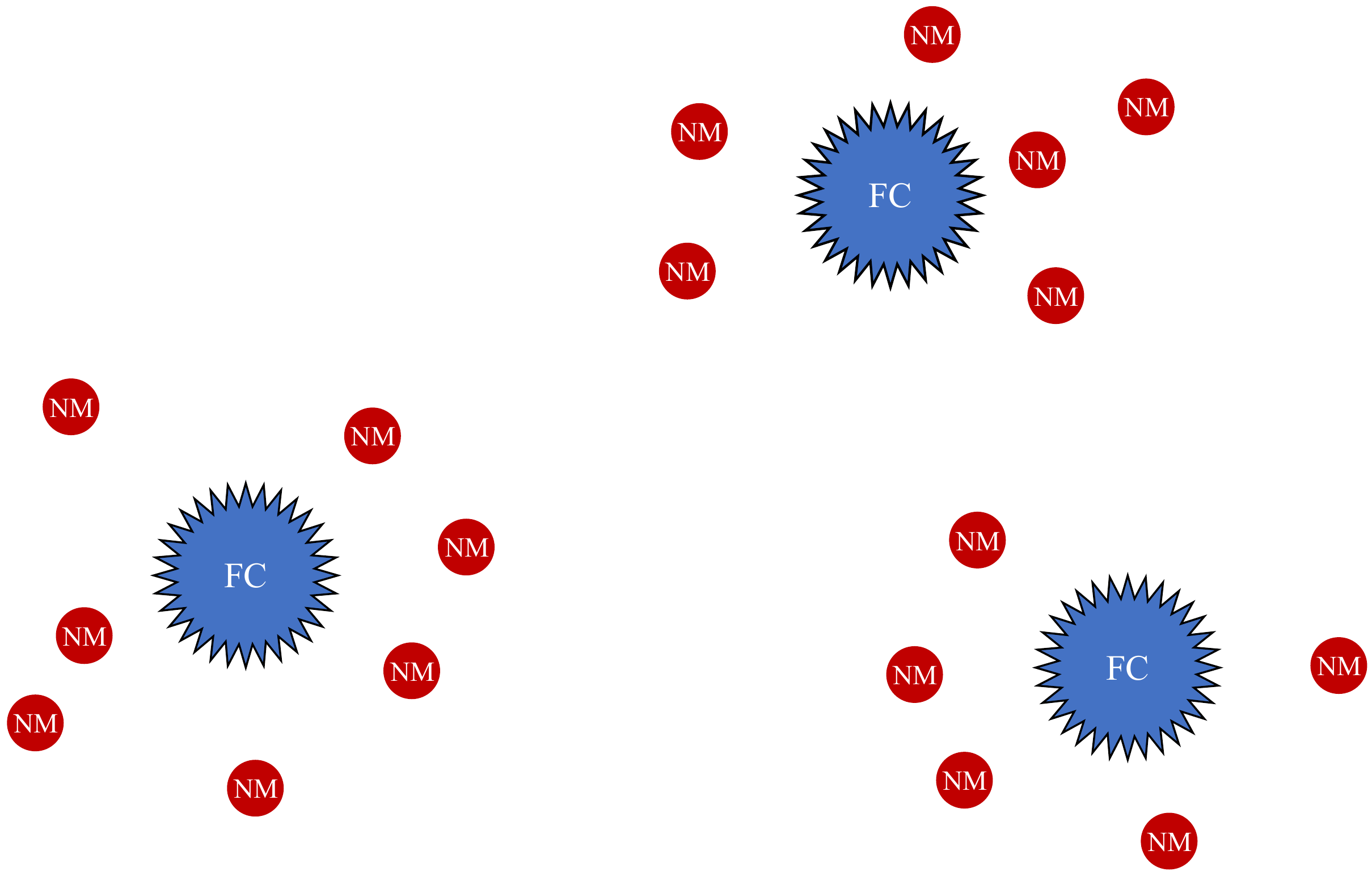}

\caption{An illustration of the system model of clustered bio-nanonetworks projected on a 2D space.}
\vspace{-10pt}
\end{figure}
\vspace{-10pt}
\subsection{Propagation Model}
We consider a time-slotted transmission with time slots of duration $T$. This implicitly assumes perfect synchronization of all NMs, which is a common assumption in the MC literature. While we will not delve into synchronization aspects in this paper, interested readers are advised to refer to \cite{clock,jamali_synch} for more information on how synchronization is achieved in MC networks.

An NM at a location $\mathbf{z}$ can release $X_\mathbf{z}$ molecules from its modulation constellation numbers $\left\{x_1,...,x_{M}\right\}$ at the beginning of a time slot to transform $\log_2(M)$ information bits, each number with equal probability $\frac{1}{M}$ \cite{modulation}. All NMs are assumed to release the same type of molecule. The molecules released by the NMs propagate through the biological medium and are observed at the FCs. NMs to FC propagation in a reference cluster is illustrated in Fig. 2. Let $p_{iL}(d)$ denote \textit{observation probability}, the probability of the event that the molecule with distance $d$ to the FC and released at the $i$-th time slot is observed in the FC at the end of the $L$-th time slot. The observation probability depends on the medium and type of detector \cite{arjmandi_main, jamali_channel}. For instance, for a general 3D environment with a point-source NM and a spherical passive FC, also referred to as transparent FC, with radius $r_0$ \footnote{The size of NMs is ignored in comparison with the size of FCs.}, the observation probabilities are given by \cite{elkas_sg_mc, jamali_channel}
\begin{align}
\label{ISI_channel}
p_{iL}(d) = g((L-i+1)T,d), \forall i \in \left\{1,..., L-1\right\},
\end{align}
and
\begin{align}
\label{main_channel}
p_{LL}(d) = g(T,d),
\end{align}
where
\begin{align}
\label{channel_function}
g(t,d) &= e^{-\mu t}\Biggl(\frac{1}{2}\left( \text{erf}\left(\frac{r_0 - d}{2\sqrt{Dt}}\right)+\text{erf}\left(\frac{r_0 + d}{2\sqrt{Dt}}\right) \right) +\nonumber\\ &\frac{\sqrt{Dt}}{\sqrt{\pi}d}\left(\exp\left(-\frac{(r_0+d)^2}{4Dt}\right) - \exp\left(-\frac{(r_0-d)^2}{4Dt}\right)\right) \Biggr),
\end{align}
where $D$ is the diffusion coefficient, which depends on the temperature, viscosity and size of the transmitted molecule. The parameter $\mu$ denotes the reaction rate constant
of molecular degradation in the environment, and $\text{erf}(x) =\frac{2}{\sqrt{\pi}}\int_{0}^{x}e^{-t^2}\mathrm{d}t$. 

In general, sequence-based channel estimators can be trained to estimate the observation probability \cite{jamali_channel}. Before we conclude this discussion, please note that the proposed spatial model is not limited to the choice of this propagation model (which is a popular one) and can be easily applied to other biological communications by choosing appropriate propagation models for those settings. 

\vspace{-8pt}

\subsection{Transmission Scheme}
In order to mitigate interference in each cluster of our clustered MC setup, in each time slot, only one of the NMs of the cluster is scheduled to release molecules according to its information availability status\footnote{The focus of this work is to answer these questions: 1) How can clustered NM transmissions be modeled? and 2) How can the resulted interference be utilized for detector analysis and design? Advanced biological models and ISI mitigation techniques are not in the scope of this work.}. Let the NM at $\mathbf{y}_\mathbf{x}^i$ in the cluster with $\mathbf{x}\in \Phi_\text{p}$ is selected to release $X_{\mathbf{x}+\mathbf{y}_\mathbf{x}^i}$ molecules in the $i$-th time slot. According to the propagation model and due to the fact that the input-output relationships between the NMs and the FCs can be modeled as independent Poisson channels with additive Poisson noise
\cite{arjmandi_main, reza_main}, at the end of $L$-th time slot, we can model the number of received molecules at the FC at $\mathbf{z}\in \Phi_\text{p}$ as
\begin{eqnarray}
\label{signal}
Y_L^{\mathbf{z}} \sim \text{Poisson} \left(\lambda_0 T+\sum_{\mathbf{x}\in\Phi_\text{p}}^{}\sum_{i=1}^{L}p_{iL}(\|\mathbf{x}+\mathbf{y}_\mathbf{x}^i\|)X_{\mathbf{x}+\mathbf{y}_\mathbf{x}^i}\right),
\end{eqnarray}
where $\lambda_0$ is the mean of the number of noise molecules in a one-second period. In \eqref{signal}, the effect of ISI due to the transmissions in the previous $i \in \left\{1,...,L-1\right\}$-th time slots is also included. 
\begin{figure}[tb!]
	\centering
	
	\includegraphics[width =3.0in]{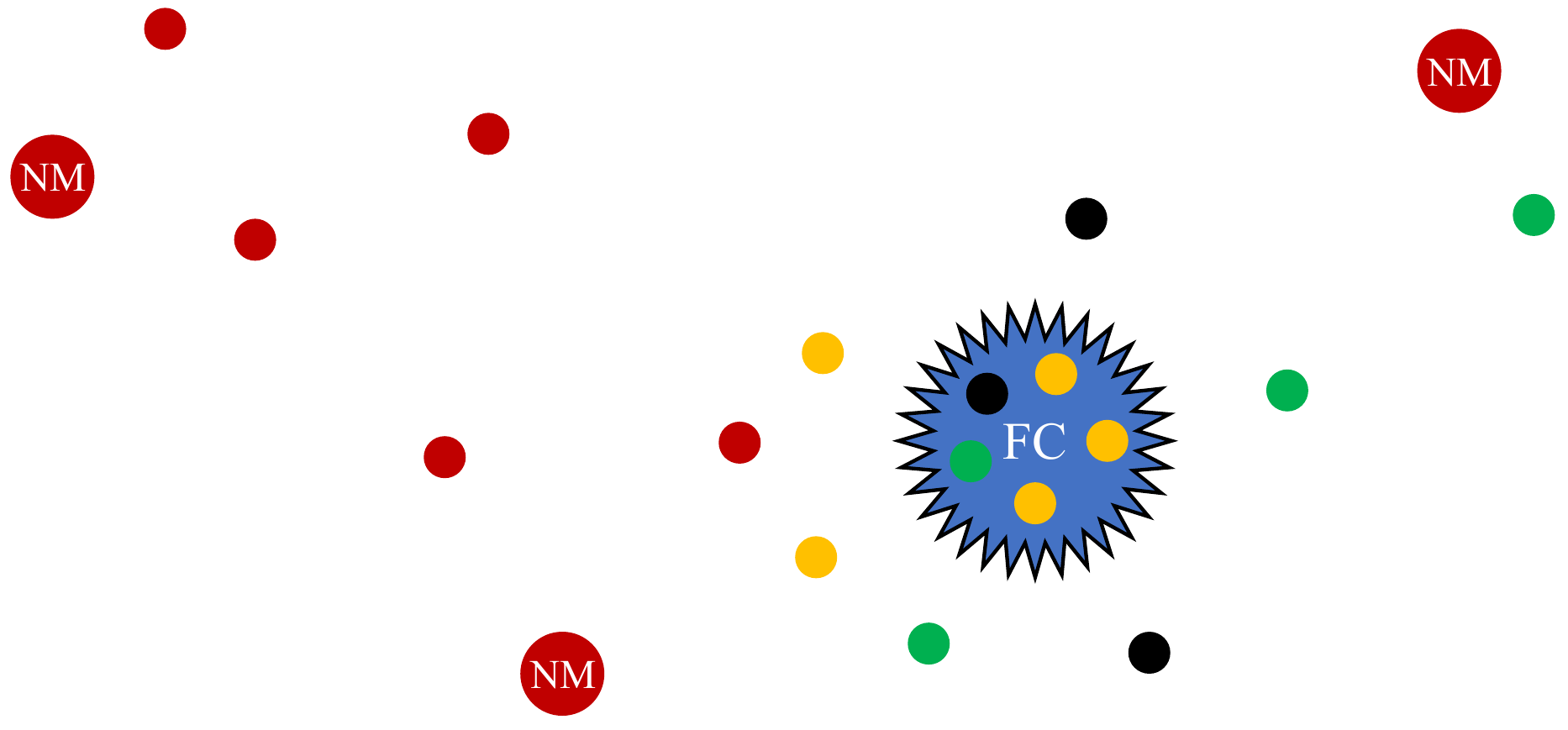}
	
	\caption{An illustrative snapshot of a reference cluster during the third time slot where the transmitted molecules in the first, second, and third time slots are shown in orange, green, and red colors, respectively. Also, noise molecules are shown in black color.}
	\vspace{-10pt}
\end{figure}

\section{Distance Distributions}
In this section, we present the following theorem on the conditional PDF of the distance of any (arbitrary) element in the set ${\cal N}^{{\mathbf{x}}}$ of the cluster centered at $\mathbf{x}\in \Phi_{\text{p}}$ to the origin $\mathbf{o}$ and the subsequent corollary as a special case of the theorem for ${\cal N}^\mathbf{o}$. These results will be used later in derivations of the expected value of interference and its Laplace transform in Section IV. As noted before already, even though the formalism for establishing distance distributions in PCPs is well-known and the distance distributions for 2D PCPs have been extensively studied in the recent years (inspired by wireless applications), similar investigations for 3D PCPs have not been performed yet. For instance, in the results below, we demonstrate that the aforementioned distribution admits a closed form solution for the 3D TCP, which was primarily enabled by the 3D assumption.
\begin{theorem} 
Conditioned on $\|\mathbf{x}\|$, i.e., the distance of the parent point $\mathbf{x}$ from the origin, the PDF of the distances $d = \|\mathbf{y}+\mathbf{x}\|, \forall \mathbf{y}\in {\cal N}^\mathbf{x}$, is
\begin{align}
\label{PCP_PDF}
&f_d(y|{\|\mathbf{x}\|}) = \int_{z_1=-y}^{y}\int_{z_2 = -\sqrt{y^2-z_1^2}}^{\sqrt{y^2-z_1^2}}\frac{y}{\sqrt{y^2-z_1^2-z_2^2}}\nonumber\\&\times\biggl[f_{\mathbf{Y}}\left(z_1-\|\mathbf{x}\|,z_2,\sqrt{y^2-z_1^2-z_2^2}\right)\nonumber\\&+f_{\mathbf{Y}}\left(z_1-\|\mathbf{x}\|,z_2,-\sqrt{y^2-z_1^2-z_2^2}\right)\biggr]\mathrm{d}z_2\mathrm{d}z_1; \ y\geq 0,
\end{align}
which is specialized for TCP as
\begin{align}
f_d(y|{\|\mathbf{x}\|}) &= \frac{y}{\sqrt{2\pi}\sigma \|\mathbf{x}\|}\biggl[\exp\left(-\frac{(y-\|\mathbf{x}\|)^2}{2\sigma^2}\right)\nonumber\\&-\exp\left(-\frac{(y+\|\mathbf{x}\|)^2}{2\sigma^2}\right)\biggr]; \ y\geq 0.
\end{align}

\end{theorem}
\begin{IEEEproof}
See Appendix A. 
\end{IEEEproof}
\begin{corollary}
The PDF of the distances $d = \|\mathbf{y}\|, \forall \mathbf{y}\in {\cal N}^\mathbf{o}$, is
\begin{align}
&f_d(y|0) = \int_{z_1=-y}^{y}\int_{z_2 = -\sqrt{y^2-z_1^2}}^{\sqrt{y^2-z_1^2}}\frac{y}{\sqrt{y^2-z_1^2-z_2^2}}\nonumber\\&\times\biggl[f_{\mathbf{Y}}\left(z_1,z_2,\sqrt{y^2-z_1^2-z_2^2}\right)\nonumber\\&+f_{\mathbf{Y}}\left(z_1,z_2,-\sqrt{y^2-z_1^2-z_2^2}\right)\biggr]\mathrm{d}z_2\mathrm{d}z_1; \ y\geq 0,
\end{align}
which is specialized for TCP as
\begin{eqnarray}
f_d(y|0) =\sqrt{\frac{2}{\pi}} \frac{y^2}{\sigma^3}\exp\left(-\frac{y^2}{2\sigma^2}\right); \ y\geq 0.
\end{eqnarray}
\end{corollary}
\begin{IEEEproof}
By putting $\|\mathbf{x}\|=0$ in Theorem 1, we get the results. However, for the TCP result, we achieve $\frac{0}{0}$ and need to use L'Hopital's rule as
\begin{align}
&\lim_{\|\mathbf{x}\|\to 0}\frac{y}{\sqrt{2\pi}\sigma \|\mathbf{x}\|}\biggl[\exp\left(-\frac{(y-\|\mathbf{x}\|)^2}{2\sigma^2}\right)\nonumber\\&-\exp\left(-\frac{(y+\|\mathbf{x}\|)^2}{2\sigma^2}\right)\biggr] = \lim_{\|\mathbf{x}\| \to 0} y \times \nonumber\\
&\frac{ \frac{y-\|\mathbf{x}\|}{\sigma^2}\exp\left(-\frac{(y-\|\mathbf{x}\|)^2}{2\sigma^2}\right)+\frac{y+\|\mathbf{x}\|}{\sigma^2}\exp\left(-\frac{(y+\|\mathbf{x}\|)^2}{2\sigma^2}\right)}{\sqrt{2\pi}\sigma} \nonumber\\
&=\sqrt{\frac{2}{\pi}} \frac{y^2}{\sigma^3}\exp\left(-\frac{y^2}{2\sigma^2}\right).
\end{align}
This completes the proof.
\end{IEEEproof}
\section{Fusion Center Detector Design}
Our analysis will focus on the performance of the typical cluster/FC, which we term as the reference cluster/FC in this paper. The conditioning of this FC at a specific location can be made rigorous by using the idea of Palm distribution. For the PPPs, it suffices to invoke Slivnyak's theorem one of whose consequences is that conditioning on the presence of a point of a homogeneous PPP at some location is equivalent to adding a point to this PPP at that location (which we assume origin without loss of generality, due to stationarity) \cite{haenggi_book}. Therefore, thanks to Slivnyak's theorem, we simply add the reference FC to the origin along with its cluster $\mathcal{N}^{\mathbf{o}}$. Also, we consider the $L$-th time slot, where the reference FC has the objective to detect the information of a reference NM at location $\mathbf{y}_0$ in the reference cluster. Then, from \eqref{signal}, the number of received molecules at the reference FC can be rewritten as
\begin{eqnarray}
Y_L^{\mathbf{o}} \sim \text{Poisson} \biggl(p_{LL}(\|\mathbf{y}_0\|) X_{\mathbf{y}_0}+{\cal I}_L+\lambda_0 T\biggr),
\end{eqnarray}
where ${\cal I}_L={\cal I}_L^{\text{intra}}+{\cal I}_L^{\text{inter}}$ is the total interference. The intra-cluster interference ${\cal I}_L^{\text{intra}}$ from NMs inside the reference cluster and inter-cluster interference ${\cal I}_L^{\text{inter}}$ from NMs of other clusters are given by
\begin{eqnarray}
{\cal I}_L^{\text{intra}} = \sum_{i=1}^{L-1}p_{iL}(\|\mathbf{y}_\mathbf{o}^i\|)X_{\mathbf{y}_\mathbf{o}^i},
\end{eqnarray}
and
\begin{eqnarray}
{\cal I}_L^{\text{inter}} = \sum_{\mathbf{x}\in\Phi_\text{p}}^{}\sum_{i=1}^{L}p_{iL}(\|\mathbf{x}+\mathbf{y}_\mathbf{x}^i\|)X_{\mathbf{x}+\mathbf{y}_\mathbf{x}^i}.
\end{eqnarray}

Due to the unknown transmitted symbols and locations of NMs, the value of ${{\cal I}_L}$ is unknown to the FCs. However, using the statistical characteristics of the interference ${{\cal I}_L}$, we propose an approximate maximum likelihood (ML) decision rule as
\begin{align}
\label{map_rule}
&\hat X_{\mathbf{y}_0} =\arg \max_{j}\mathbb{P}(Y_L^{\mathbf{o}} = y \mid X_{\mathbf{y}_0}=x_j) \approx \arg \max_{j}\nonumber\\& \frac{e^{-\left(p_{LL}(\|\mathbf{y}_0\|) x_j+\mathbb{E}\left\{{{\cal I}_L}\right\}+\lambda_0 T\right)}(p_{LL}(\|\mathbf{y}_0\|) x_j+\mathbb{E}\left\{{{\cal I}_L}\right\}+\lambda_0 T)^y}{y!},
\end{align}
where the interference ${{\cal I}_L}$ is approximated by its expected value $\mathbb{E}\left\{{\cal I}_L\right\}$, given in the following lemma.
\begin{lemma}
The expected value of interference ${\cal I}_L$ is
\begin{eqnarray}
\label{total_int}
\mathbb{E}\left\{{\cal I}_L\right\} = \mathbb{E}\left\{{\cal I}_L^{\text{intra}}\right\} + \mathbb{E}\left\{{\cal I}_L^{\text{inter}}\right\},
\end{eqnarray}
where
\begin{align}
\label{intra_cluster_omid2}
&\mathbb{E}\left\{{\cal I}_L^{\text{intra}}\right\} =\nonumber\\& \frac{\sum_{j=1}^{M}{x_j}}{M\int_{r_0}^{\infty}f_{d}(y|0)\mathrm{d}y}\int_{r_0}^{\infty}f_{d}(y|0)\sum_{i=1}^{L-1}p_{iL}(y)\mathrm{d}y,
\end{align}
and
\begin{align}
\label{inter_cluster_omid2}
\mathbb{E}\left\{{\cal I}_L^{\text{inter}}\right\} &= \frac{4\pi \lambda_\text{p}\sum_{j=1}^{M}{x_j}}{M} \times \nonumber\\&\int_{2r_0}^{\infty} \frac{x^2\int_{r_0}^{\infty} f_{d}(y|x)\sum_{i=1}^{L}p_{iL}(y)\mathrm{d}y}{\int_{r0}^{\infty}f_d(y|x)\mathrm{d}y} \mathrm{d}x.
\end{align}
\end{lemma}
\begin{IEEEproof}
	See Appendix B.
\end{IEEEproof}
From \eqref{inter_cluster_omid2}, it can be seen that the expected interference linearly increases with $\lambda_\text{p}$ and $\frac{1}{M}\sum_{j=1}^{M}x_j$.

Assuming the order $x_1<x_2<...<x_M$, it can be shown that the rule in \eqref{map_rule} can be further simplified to \cite{reza_main}
\begin{eqnarray}
\hat X_{\mathbf{y}_0} = \left\{\begin{matrix}
x_1 & \text{if}  \hspace{+5pt} y<\text{th}_1,\\
x_j & \hspace{40pt}\text{if}  \hspace{+5pt} \text{th}_{j-1}\leq y<\text{th}_j,\\
\hspace{3pt}x_M & \hspace{14pt}\text{if}  \hspace{+5pt} y\geq \text{th}_{M-1},\\
 \end{matrix}\right.
\end{eqnarray}
where the thresholds are as
\begin{align}
\label{threshold}
\text{th}_j = \Biggl \lceil\frac{p_{LL}(\|\mathbf{y}_0\|) {x_{j+1}}-p_{LL}(\|\mathbf{y}_0\|) {x_{j}}}{{\ln}\left(\frac{p_{LL}(\|\mathbf{y}_0\|) {x_{j+1}}+\mathbb{E}\left\{{{\cal I}_L}\right\}+\lambda_0 T}{p_{LL}(\|\mathbf{y}_0\|) {x_{j}}+\mathbb{E}\left\{{{\cal I}_L}\right\}+\lambda_0 T}\right)}\Biggr \rceil,\ j\in \left\{1,...,M-1\right\},
\end{align}
where $\lceil a \rceil$ denotes the closest upper integer to $a$. This simple detector has low complexity to be implemented on an NM as $\mathbb{E}\left\{{\cal I}_L\right\}, \forall L$ in \eqref{threshold} can be evaluated offline with no channel knowledge.

\section{Performance Analysis}
In this section, we derive the error probability of the detector proposed in Section IV for the considered PCP setup. In order to put the novelty of this analysis into context, please note that the error probability analysis in the literature has been limited to the PPP-based models (assuming $M=2$) \cite{elkas_sg_mc, dissan_sg_mc, gupta_sg_mc}. In fact, even the single-cluster setup (which is a special case of our setup) has not been studied yet. From the technical standpoint, we obtain a new form of the Laplace tansform of interference, propose new approaches to evaluate derivatives that appear in this analysis, as well as propose new special cases and a closed-form upper bound.

The probability of error $\cal E$ of the proposed detector can be written as
\begin{eqnarray}
\label{error}
\mathbb{P} ({\cal E}) = \frac{1}{M}\sum_{j=1}^{M}\mathbb{P}({\cal E}\mid X_{\mathbf{y}_0}=x_j),
\end{eqnarray}
where, defining $\text{th}_0 = 0$, $\text{th}_M = \infty$, and the Laplace transform (LT) of ${\cal I}_{L}$ as ${{\cal L}_{{\cal I}_{L}}}\left( s \right) = \mathbb{E}\left\{e^{-s{\cal I}_{L}}\right\}$, we have
\begin{align}
\label{BEP}
&\mathbb{P} ({\cal E}\mid X_{\mathbf{y}_0}=x_j) = \mathbb{P}(y<\text{th}_{j-1}, \ y\geq \text{th}_j \mid X_{\mathbf{y}_0}=x_j) \nonumber\\
&= 1- \mathbb{P}(\text{th}_{j-1}\leq y<\text{th}_{j} \mid X_{\mathbf{y}_0}=x_j) =1- \mathbb{E}\Biggl\{\sum_{k=\text{th}_{j-1}}^{\text{th}_{j}-1} \nonumber\\&\frac{e^{-\left(p_{LL}(\|\mathbf{y}_0\|) x_j+{{\cal I}_L}+\lambda_0 T\right)}(p_{LL}(\|\mathbf{y}_0\|) x_j+{{\cal I}_L}+\lambda_0 T)^k}{k!}\Biggr\}\nonumber\\&
=1- \sum_{k=\text{th}_{j-1}}^{\text{th}_j-1} \frac{(-1)^k}{k!} \frac{\mathrm{d}^k }{\mathrm{d} s^k}e^{-s(p_{LL}(\|\mathbf{y}_0\|) x_j+\lambda_0 T)}{\cal L}_{{\cal I}_L}(s)\bigg|_{s = 1},
\end{align}
which follows from $\mathbb{E}\left\{e^{-s \left(p_{LL}(\|\mathbf{y}_0\|) x_j+{{\cal I}_L}+\lambda_0 T\right)} \right\} = e^{-s \left(p_{LL}(\|\mathbf{y}_0\|) x_j+\lambda_0 T\right)}{{\cal L}_{{{\cal I}_L}}}\left( s \right), \forall s$, and the derivative property of the LT as $\mathbb{E}\left\{X^m e^{-sX}\right\} = (-1)^m \frac{\mathrm{d}^m {\cal L}_{X}(s)}{\mathrm{d} s^m}, \forall m$, for a random variable $X$. In \eqref{BEP}, ${{\cal L}_{{\cal I}_{L}}}$ is given in Lemma 2.

\begin{lemma}
The LT of the interference ${{\cal I}_L}$ is
\begin{eqnarray}
{\cal L}_{{\cal I}_L} (s) = {\cal L}_{{\cal I}_L^{\text{intra}}} (s) \times {\cal L}_{{\cal I}_L^{\text{inter}}} (s),
\end{eqnarray}
where
\begin{align}
&{\cal L}_{{\cal I}_L^{\text{intra}}} (s) =\nonumber\\&\frac{1}{\left(M\int_{r_0}^{\infty}  f_d(y|0)\mathrm{d}y\right)^{L-1}}\mathop \prod \limits_{i=1}^{L-1}\int_{r_0}^{\infty}\sum_{j=1}^{M} e^{-s x_j p_{iL}(y)} f_d(y|0)\mathrm{d}y,
\end{align}
and
\begin{align}
&{\cal L}_{{\cal I}_L^{\text{inter}}} (s) =\exp \Biggl(-4\pi\lambda_\text{p} \int_{2r_0}^{\infty}\biggl(1-\frac{1}{(M\int_{r0}^{\infty}f_d(y|x)\mathrm{d}y)^L}\times\nonumber\\&\mathop \prod \limits_{i=1}^{L}\int_{r_0}^{\infty}\sum_{j=1}^{M}{e^{-s x_j p_{iL}{(y)}}}f_d(y|x)\mathrm{d}y\biggr)x^2\mathrm{d}x \Biggr).
\end{align}
\end{lemma}
\begin{IEEEproof}
	See Appendix C.
\end{IEEEproof}
Then, using the general Leibniz rule \cite{lebenan}, the term $\frac{\mathrm{d}^k }{\mathrm{d} s^k}e^{-s(p_{LL}(\|\mathbf{y}_0\|) x_j+\lambda_0 T)}{\cal L}_{{\cal I}_L}(s)\bigg|_{s = 1}$ in \eqref{BEP} can be computed as
\begin{align}
\label{step1}
&\frac{\mathrm{d}^k }{\mathrm{d} s^k}e^{-s(p_{LL}(\|\mathbf{y}_0\|) x_j+\lambda_0 T)}{\cal L}_{{\cal I}_L}(s)\bigg|_{s = 1}=e^{-(p_{LL}(\|\mathbf{y}_0\|) x_j+\lambda_0 T)}\times \nonumber\\& \sum_{k_1+k_2+k_3 = k}\frac{k!}{k_1!k_2!k_3!}(-1)^{k_1} (p_{LL}(\|\mathbf{y}_0\|) x_j+\lambda_0 T)^{k_1}\times\nonumber\\&\frac{\mathrm{d}^{k_2} {\cal L}_{{\cal I}_L^{\text{intra}}} (s)}{\mathrm{d} s^{k_2}}\bigg|_{s = 1}\frac{\mathrm{d}^{k_3} {\cal L}_{{\cal I}_L^{\text{inter}}} (s)}{\mathrm{d} s^{k_3}}\bigg|_{s = 1},
\end{align}
where the sum is over all 3-tuples $(k_1, k_2, k_3)$ of non-negative integers with $k_1+k_2+k_3 = k$. In \eqref{step1}, from the general Leibniz rule, we have
\begin{align}
\label{step2}
&\frac{\mathrm{d}^{k_2} {\cal L}_{{\cal I}_L^{\text{intra}}} (s)}{\mathrm{d} s^{k_2}}\bigg|_{s = 1} = \frac{\mathrm{d}^{k_2} }{\mathrm{d} s^{k_2}}\frac{1}{\left(M\int_{r_0}^{\infty}  f_d(y|0)\mathrm{d}y\right)^{L-1}}\times\nonumber\\
&\mathop \prod \limits_{i=1}^{L-1}\int_{r_0}^{\infty}\sum_{j=1}^{M} e^{-s x_j p_{iL}(y)} f_d(y|0)\mathrm{d}y\bigg|_{s = 1}\nonumber\\
&=\frac{1}{\left(M\int_{r_0}^{\infty}  f_d(y|0)\mathrm{d}y\right)^{L-1}}\sum_{k_{2,1}+\cdots+k_{2,{L-1}}=k_2} \frac{k_2!}{k_{2,1}!...k_{2,L-1}!} \nonumber\\&
\mathop \prod \limits_{i=1}^{L-1}(-1)^{k_{2,i}} \int_{r_0}^{\infty}p_{iL}(y)^{k_{2,i}}\sum_{j=1}^{M}  x_j^{k_{2,i}} e^{- x_j p_{iL}(y)} f_d(y|0)\mathrm{d}y,
\end{align}
where the sum extends over all $L-1$-tuples $(k_{2,1}, \cdots, k_{2,L-1})$ of non-negative integers with $k_{2,1}+\cdots+k_{2,{L-1}}=k_2$. Finally, using the Fa\`a� di Bruno's formula \cite{faadi} for the last term in \eqref{step1}, we obtain
\begin{align}
&\frac{\mathrm{d}^{k_3} {\cal L}_{{\cal I}_L^{\text{inter}}} (s)}{\mathrm{d} s^{k_3}}\bigg|_{s = 1} = \frac{\mathrm{d}^{k_3} }{\mathrm{d} s^{k_3}}\exp \Biggl(-4\pi\lambda_\text{p} \int_{2r_0}^{\infty}\biggl(1-\nonumber\\&\frac{1}{(M\int_{r0}^{\infty}f_d(y|x)\mathrm{d}y)^L}\mathop \prod \limits_{i=1}^{L}\int_{r_0}^{\infty}\sum_{j=1}^{M}{e^{-s x_j p_{iL}{(y)}}}f_d(y|x)\mathrm{d}y\biggr)\nonumber\\&\times x^2\mathrm{d}x \Biggr)\bigg|_{s = 1}= \exp \Biggl(-4\pi\lambda_\text{p} \int_{2r_0}^{\infty}\biggl(1-\frac{1}{(M\int_{r0}^{\infty}f_d(y|x)\mathrm{d}y)^L}\nonumber\\&\times\mathop \prod \limits_{i=1}^{L}\int_{r_0}^{\infty}\sum_{j=1}^{M}{e^{- x_j p_{iL}{(y)}}}f_d(y|x)\mathrm{d}y\biggr)x^2\mathrm{d}x \Biggr)\times\nonumber\\
\label{step3}
&\sum_{k_3} \frac{k_3!}{k_{3,1}!1!^{k_{3,1}}\cdots k_{3,k_3}!k_3!^{k_{3,k_3}}}\times\nonumber\\
&\mathop \prod \limits_{l=1}^{k_3}\Biggl(\frac{\mathrm{d}^{l} }{\mathrm{d} s^{l}}-4\pi\lambda_\text{p} \int_{2r_0}^{\infty}\Biggl(1-\frac{1}{(M\int_{r0}^{\infty}f_d(y|x)\mathrm{d}y)^L}\times\nonumber\\&\mathop \prod \limits_{i=1}^{L}\int_{r_0}^{\infty}\sum_{j=1}^{M}{e^{-s x_j p_{iL}{(y)}}}f_d(y|x)\mathrm{d}y\Biggr)x^2\mathrm{d}x \bigg|_{s =1}\Biggr)^{k_{3,l}},
\end{align}
where the summation $\sum_{k_3}$ is over all $k_3$-tuples of nonegative integers $(k_{3,1},\cdots, k_{3,k_3})$ satisfying the constraint $1 \ .\ k_{3,1}+2 \ . \ k_{3,2}+\cdots+k_3 \ . \ k_{3,k_3} = k_3$. In \eqref{step3}, the $l$-th order derivative term is calculated as
\begin{align}
\label{final_formula}
&\frac{\mathrm{d}^{l} }{\mathrm{d} s^{l}}-4\pi\lambda_\text{p} \int_{2r_0}^{\infty}\Biggl(1-\frac{1}{(M\int_{r0}^{\infty}f_d(y|x)\mathrm{d}y)^L}\mathop \prod \limits_{i=1}^{L}\int_{r_0}^{\infty}\nonumber\\&\sum_{j=1}^{M}{e^{-s x_j p_{iL}{(y)}}}f_d(y|x)\mathrm{d}y\Biggr)x^2\mathrm{d}x \bigg|_{s =1}=4\pi\lambda_\text{p}\times\nonumber\\& \int_{2r_0}^{\infty}\frac{1}{(M\int_{r0}^{\infty}f_d(y|x)\mathrm{d}y)^L}\frac{\mathrm{d}^{l} }{\mathrm{d} s^{l}}\mathop \prod \limits_{i=1}^{L}\int_{r_0}^{\infty}\sum_{j=1}^{M}{e^{-s x_j p_{iL}{(y)}}}\nonumber\\&\times f_d(y|x)\mathrm{d}yx^2\mathrm{d}x\bigg|_{s =1}=4\pi\lambda_\text{p}\int_{2r_0}^{\infty}\frac{1}{\left(M\int_{r_0}^{\infty}  f_d(y|x)\mathrm{d}y\right)^{L}}\nonumber
\end{align}
\begin{align}
&\times \sum_{l_{1}+\cdots+l_{{L}}=l} \frac{l!}{l_{1}!...l_{L}!}  \mathop \prod \limits_{i=1}^{L}(-1)^{l_{i}}\times\nonumber\\& \int_{r_0}^{\infty}p_{iL}(y)^{l_{i}}\sum_{j=1}^{M}  x_j^{l_{i}} e^{- x_j p_{iL}(y)} f_d(y|x)\mathrm{d}yx^2\mathrm{d}x,
\end{align}
where the sum is taken over all $L$-tuples $(l_{1}, \cdots, l_{L})$ of non-negative integers with $l_{1}+\cdots+l_{{L}}=l$.

As the form given in \eqref{BEP} is rather unwieldy, we next provide a closed-form upper-bound for \eqref{BEP}.

\begin{theorem}
The conditional probability of error \eqref{BEP} can be upper-bounded by
\begin{align}
\label{upper_1}
&\mathbb{P}_\text{u} ({\cal E}\mid X_{\mathbf{y}_0}=x_j) = 1- \sum_{k=1}^{{\text{th}}_j} (-1)^{k+1} {{{\text{th}}_j} \choose {k}} \times \nonumber\\&e^{-{\eta}_j (p_{LL}(\|\mathbf{y}_0\|) x_j+\lambda_0 T) k}{\cal L}_{{\cal I}_L}({\eta}_jk) + \sum_{k=1}^{{\text{th}}_{j-1}} (-1)^{k+1} {{{\text{th}}_{j-1}} \choose {k}} \nonumber\\&\times e^{-{\eta}_{j-1} (p_{LL}(\|\mathbf{y}_0\|) x_j+\lambda_0 T) k}{\cal L}_{{\cal I}_L}({\eta}_{j-1}k), j\in \left\{1,...,M-1\right\},
\end{align}
and
\begin{align}
\label{upper_2}
&\mathbb{P}_\text{u} ({\cal E}\mid X_{\mathbf{y}_0}=x_M) =  \sum_{k=1}^{{\text{th}}_{M-1}} (-1)^{k+1} {{{\text{th}}_{M-1}} \choose {k}} \nonumber\\&\times e^{-{\eta}_{M-1} (p_{LL}(\|\mathbf{y}_0\|) x_M+\lambda_0 T) k}{\cal L}_{{\cal I}_L}({\eta}_{M-1}k),
\end{align}
where ${\eta}_{j} = {\text{th}_{j}!}^{-\frac{1}{\text{th}_{j}}}, j\in \left\{1,...,M-1\right\}$, and ${\cal L}_{{\cal I}_L}$ is given in Lemma 2.
\end{theorem}
\begin{IEEEproof}
See Appendix D.
\end{IEEEproof}

It is notable that this upper-bound can be also used for other setups when we replace related LT of interference. In the following lemma, the design and analysis results are simplifed for the ON/OFF keying modulation \cite{reza_main, elkas_sg_mc, gupta_sg_mc, modulation}.
\begin{lemma} 
In the special case $M=2$ with $x_0 = 0$ and $x_1 = \xi$, when 
\begin{align}
\label{xi}
\xi &< - \frac{\mathbb{E}\left\{{\cal I}_{L}\right\}+\lambda_0 T}{p_{LL}(\|\mathbf{y}_0\|)} -\frac{1}{p_{LL}(\|\mathbf{y}_0\|)} {\cal W}\bigl(-({\mathbb{E}\left\{{\cal I}_{L}\right\}+\lambda_0 T})\nonumber\\&\times\exp\left(-{\mathbb{E}\left\{{\cal I}_{L}\right\}+\lambda_0 T}\right)\bigr),
\end{align}
where $\cal W$ is the Lambert function \cite{lambert}, the error probability is simplified to
\begin{align}
\mathbb{P} ({\cal E}) = \frac{1}{2}\mathbb{P} ({\cal E}|x_0)+ \frac{1}{2}\mathbb{P} ({\cal E}|x_1),
\end{align}
where
\begin{align}
\mathbb{P} ({\cal E}|x_0) = 1 - e^{-\lambda_0 T} {\cal L}_{{\cal I}_L}(1),
\end{align}
and
\begin{align}
\mathbb{P} ({\cal E}|x_1) = e^{-\left(p_{LL}(\|\mathbf{y}_0\|\right)\xi+\lambda_0 T)} {\cal L}_{{\cal I}_L}(1).
\end{align}

\end{lemma}
\begin{IEEEproof}
See Appendix E.
\end{IEEEproof}
\section{Numerical Results}
In this section, we provide numerical results for specific scenarios of clustered bio-nanonetworks with the parameter values given in Table 1, unless otherwise
stated. We also provide Monte Carlo simulations with 50000 realizations to verify the accuracy of the results. For the simulations, we consider the biological environment as $[-25r_0,25r_0]$ for each dimension and include exclusion zones for all the FCs to exactly model the clustered setup described in Section II.A. Recall that the exclusion zone for the reference FC was incorporated in the analysis too. Therefore, this true simulation and our analysis differ in terms of exclusion zones for the interfering FCs, which were considered in the simulation but ignored in the analysis. However, numerical comparisons of these two cases will shortly reveal that ignoring exclusion zones of the interfering FCs in the analysis was without any loss of accuracy. Also, while we presented the analytical results for a general $M$, here we focus on $M = 2$ as it is the usual present implementation case for molecular communications \cite{reza_main, akan_sg_mc, elkas_sg_mc, dissan_sg_mc, gupta_sg_mc, modulation} and the effect of different parameters can be described more cleanly. 

In the following, the analytical results are calculated from \eqref{error}-\eqref{final_formula} and the upper-bound results are from \eqref{error}, \eqref{upper_1}-\eqref{upper_2}.

In Fig. 3, the analytical results, upper-bound results, and Monte Carlo simulations for the error probability are shown as a function of the number of time slot $L$ for $D = 10 \times 10^{-12}\ \frac{\text{m}^2}{\text{s}}$ and $40 \times 10^{-12}\ \frac{\text{m}^2}{\text{s}}$. It is observed that the analytical and upper-bound results tightly mimic the exact Monte Carlo simulations. Also, the error probability increases as the number of time slots increases. It is because the interference from ISI increases. However, for $L>5$, the increase is slight. This is from the fact that the observation probabilities $p_{iL}$ in (2) from previous $ \left\{1,...,L-5\right\}$-th time slots can be ignored. Other observation is related to the increase of the diffusion coefficient $D$ that can significantly decrease the error probability. As per equation (3), an increase in $D$ from $10 \times 10^{-12}\ \frac{\text{m}^2}{\text{s}}$ to $40 \times 10^{-12}\ \frac{\text{m}^2}{\text{s}}$ leads to a rise in the observation probability $p_{LL}$, resulting in an improvement in the link from the reference NM. However, this increase in $D$ also causes interference to increase. Despite this, the first one remains the dominant factor.

The error probability as a function of time slot duration $T$ for $x_2 = 120$ and $\|\mathbf{y}_0\| = 3.5r_0$ and $4r_0$ is studied in Fig. 4. It is observed that, depending on $\|\mathbf{y}_0\|$, there is an optimal
value for $T$, about $1.3\ \text{s}$ for $\|\mathbf{y}_0\| = 3.5r_0$ and $1.7 \ \text{s}$ for $\|\mathbf{y}_0\| = 4r_0$. This shows a tradeoff. As $T$ increases, the observation probabilities $p_{iL}$ in (2) decrease, which decreases the ISI. On the other hand, the observation probability $p_{LL}$ in (3) from the reference NM decreases. The first factor dominates the error probability in a lower $T$, and for a higher $T$, the second factor plays a more important role. Also, when the reference NM is located closer to the reference FC, the performance improves since $p_{LL}$ increases. In Fig. 5, the expected number of interference molecules in \eqref{total_int} and the threshold given in \eqref{threshold} are shown as a function of $T$. We can observe that as $T$ increases, the expected number of interference molecules decreases and the threshold follows the decreasing trend by unit step function decrements. This decrease is because of the decrease in ISI as more molecules from the previous time slots leave the FC after a higher $T$.

In Fig. 6, the error probability as a function of the parent intensity $\lambda_\text{p}$ is plotted for $\sigma = 3r_0$ and $5r_0$. The error probability increases with the increase in $\lambda_\text{p}$. It is because the number of interfering clusters increases. Also, increasing $\sigma$ improves the performance. This is intuitive because the probability of the event that the reference FC is located farther from the NMs of the reference cluster increases with $\sigma$ and this leads to decrease in $p_{iL}$ in (2). In Fig. 7, the expected number of interference and the threshold are shown as a function of $\lambda_\text{p}$ for $\sigma = 3r_0$. We observe that the number of interference molecules increases linearly with $\lambda_\text{p}$ and the threshold follows this increase. 

Figs. 8 and 9 present the effect of the number of released molecules for the second symbol, i.e., $x_2$, on the error probability for $\mu = 0.1\ \text{s}^{-1}$ and $0.8 \ \text{s}^{-1}$, and on the expected number of interference and the threshold for $\mu = 0.1 \ \text{s}^{-1}$, respectively. It is observed that by increasing $x_2$ (increasing $x_2-x_1$), the error probability decreases. This is because from \eqref{threshold}, the decision area for each symbol expands. However, the slope of this improvement in performance decreases, which is due to the increase in interference. Furthermore, an increase in $\mu$ results in a reduction in the error probability as it lowers both the observed interference and the desired molecules. However, in the case of typical $\mu$ values, the decrease in interference has a more significant impact. It is also observed that the threshold can adjust itself with the increase in $x_2$.  
\begin{table}[t]
	\caption {Parameter Values} 
	\vspace{-8pt}
	\begin{center}
		\resizebox{9.2cm}{!} {
			\begin{tabular}{| l | l | l | l | l | l | l | l | l | l | l | l | l | l | l |}
				
				\hline
				\hline
				{$\lambda_{\rm p}$}&{$r_0$}&$\|\mathbf{y}_0\|$&{$D$}& $\mu$& $\sigma$& $T$& $L$& $M$ & $\left\{x_1,x_2\right\}$ \\ \hline
				$2 \times 10^{-6}$ $\mu\text{m}^{3}$& 5 $\mu\text{m}$&$2r_0$&$40\times 10^{-12}$ $\frac{\text{m}^2}{\text{s}}$ & 0.1 $\text{s}^{-1}$& 20 $\mu \text{m}$ & 0.5 $\text{s}$ & 5 & 2 & $\left\{0, 60\right\}$\\ \hline
				
				\hline
		\end{tabular}}
		
	\end{center}
	\vspace{-7pt}
\end{table}

\begin{figure}[tb!]
	\vspace{0pt}
	\centering
	\includegraphics[width =3.2in]{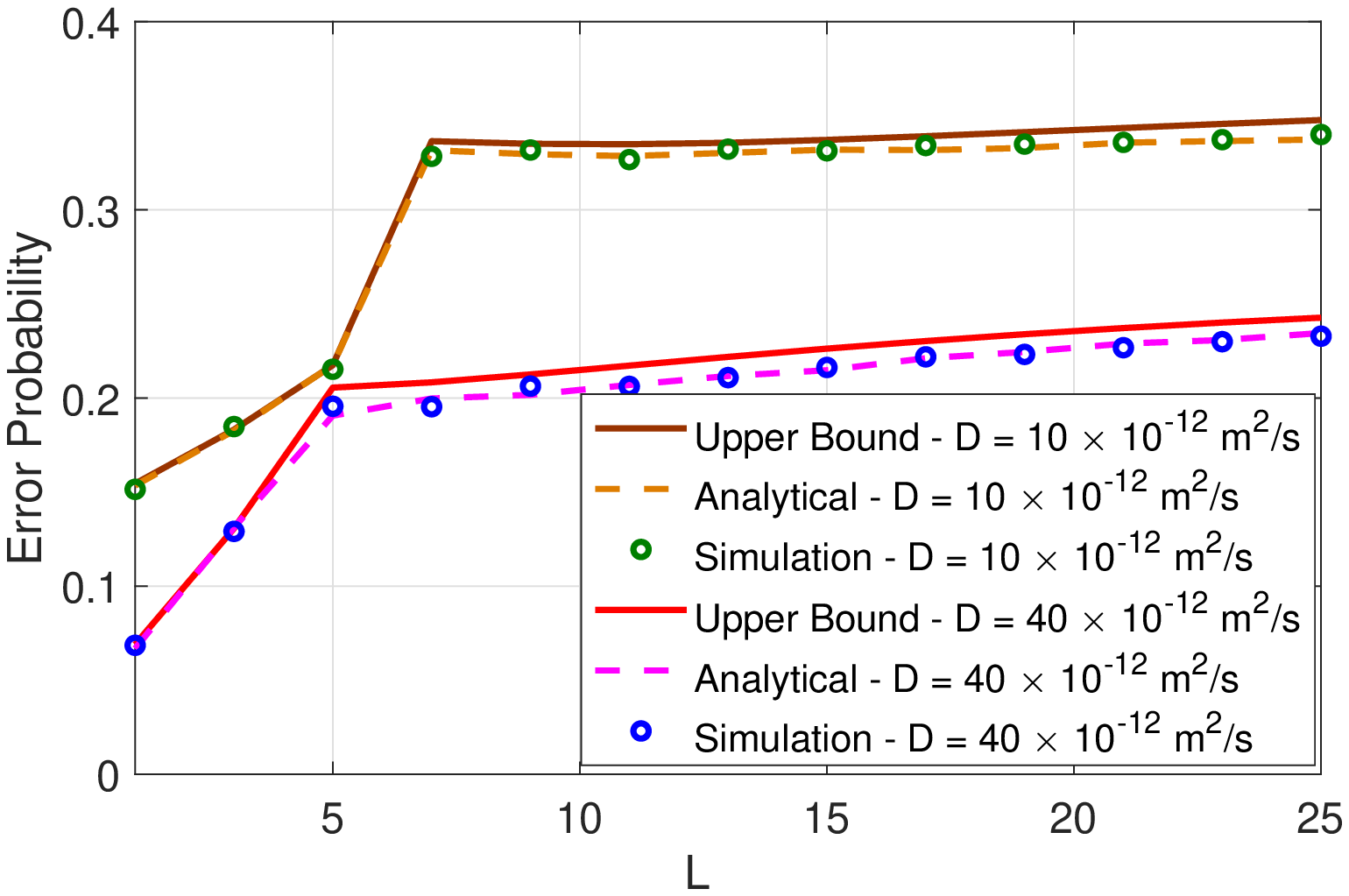} 
	\vspace{-3pt}
	\caption{Error probability as a function of $L$.}
	\vspace{-8pt}
\end{figure}

\begin{figure}[tb!]
	\vspace{0pt}
	\centering
	\includegraphics[width =3.2in]{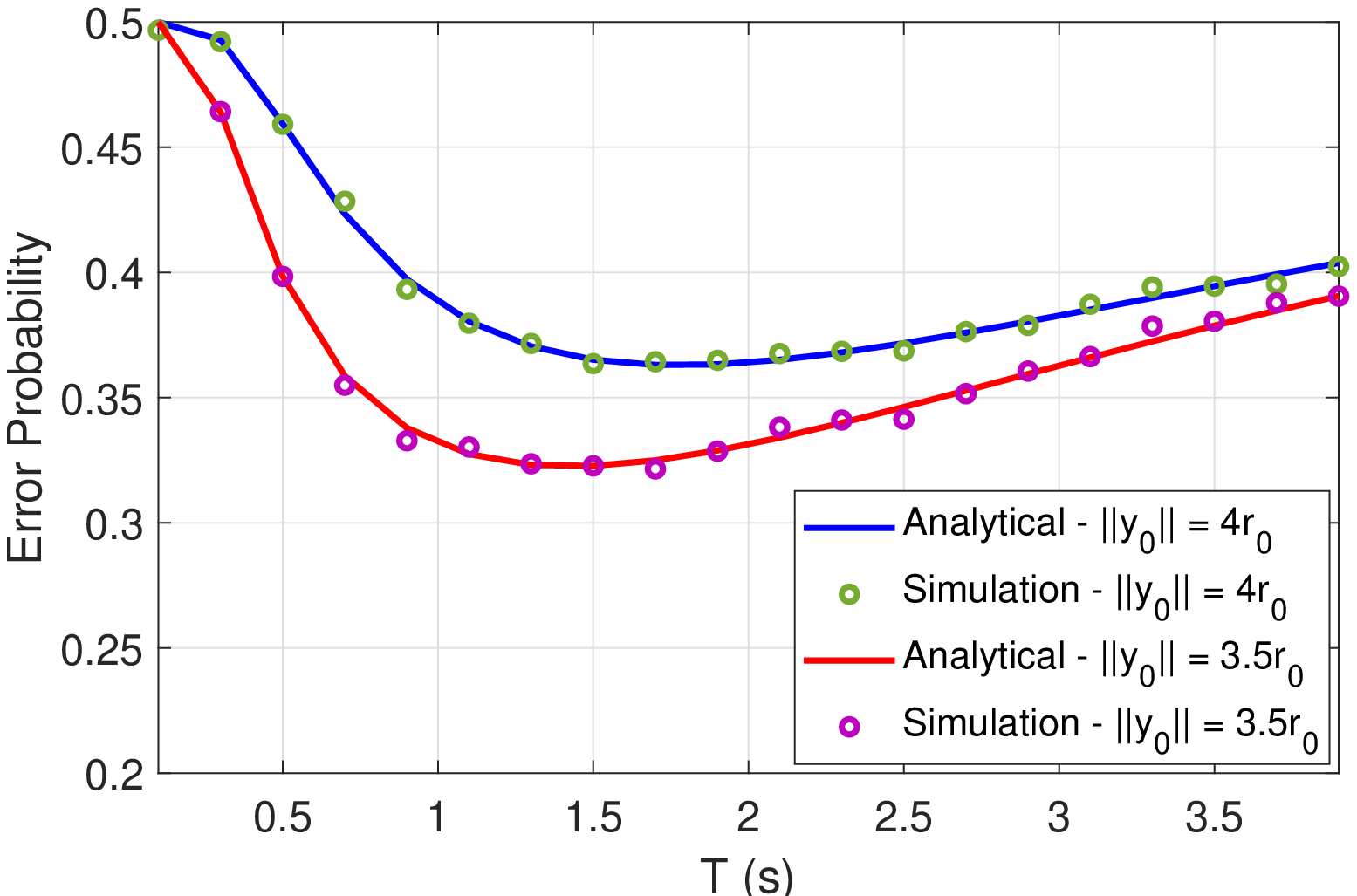} 
	\vspace{-3pt}
	\caption{Error probability as a function of $T$.}
	\vspace{-8pt}
\end{figure}

\begin{figure}[tb!]
	\vspace{0pt}
	\centering
	\includegraphics[width =3.2in]{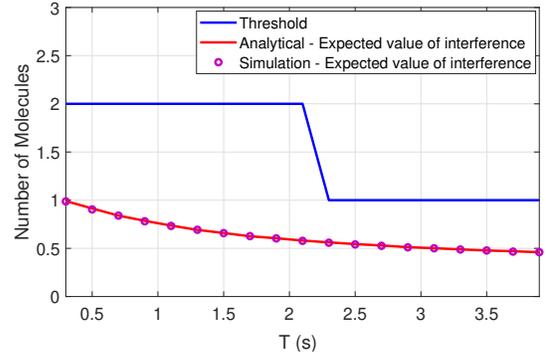} 
	\vspace{-3pt}
	\caption{Number of molecules as a function of $T$.}
	\vspace{-8pt}
\end{figure}

\begin{figure}[tb!]
	\vspace{0pt}
	\centering
	\includegraphics[width =3.2in]{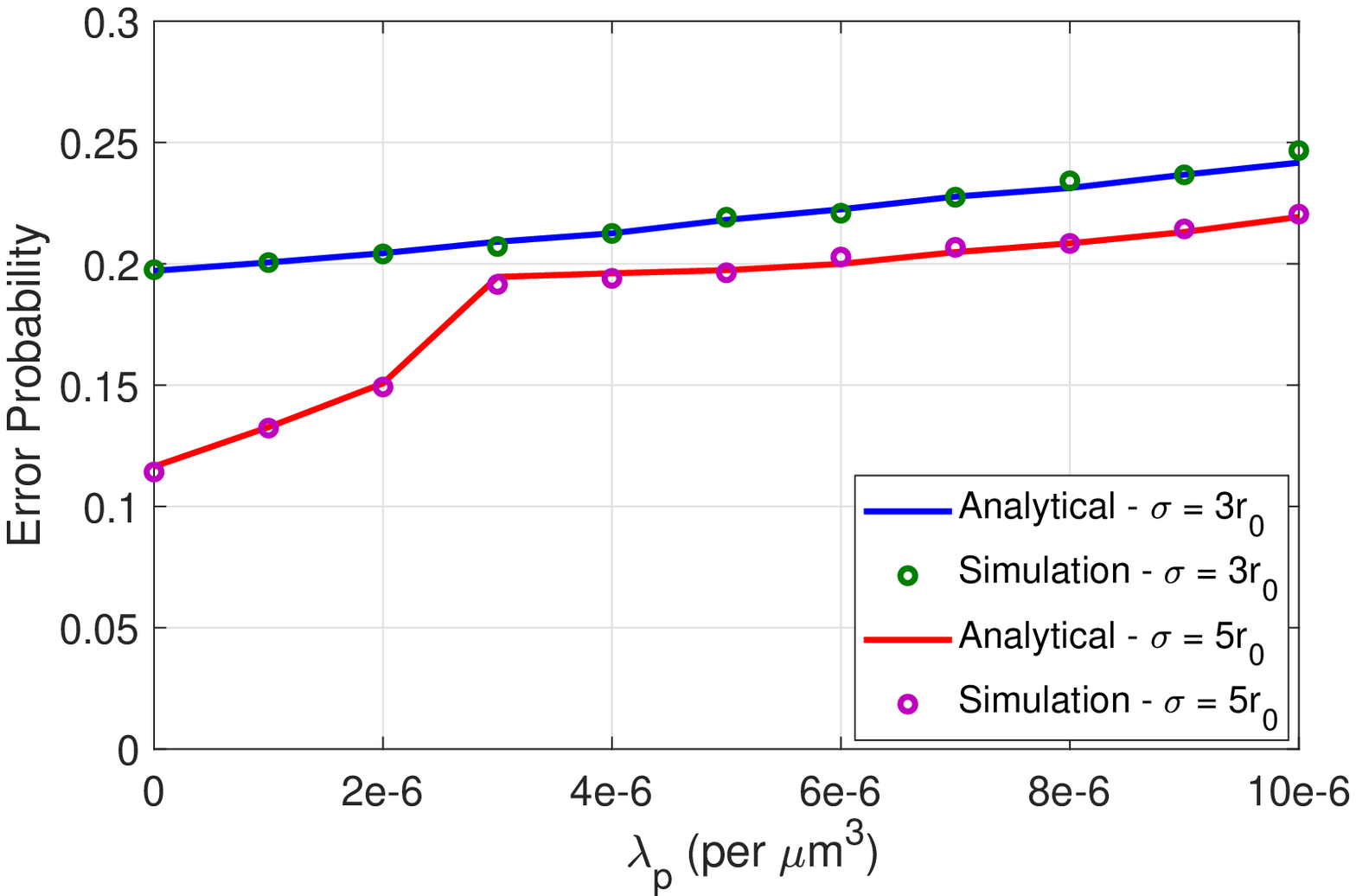} 
	\vspace{-3pt}
	\caption{Error probability as a function of $\lambda_\text{p}$.}
	\vspace{-8pt}
\end{figure}

\begin{figure}[tb!]
	\vspace{0pt}
	\centering
	\includegraphics[width =3.2in]{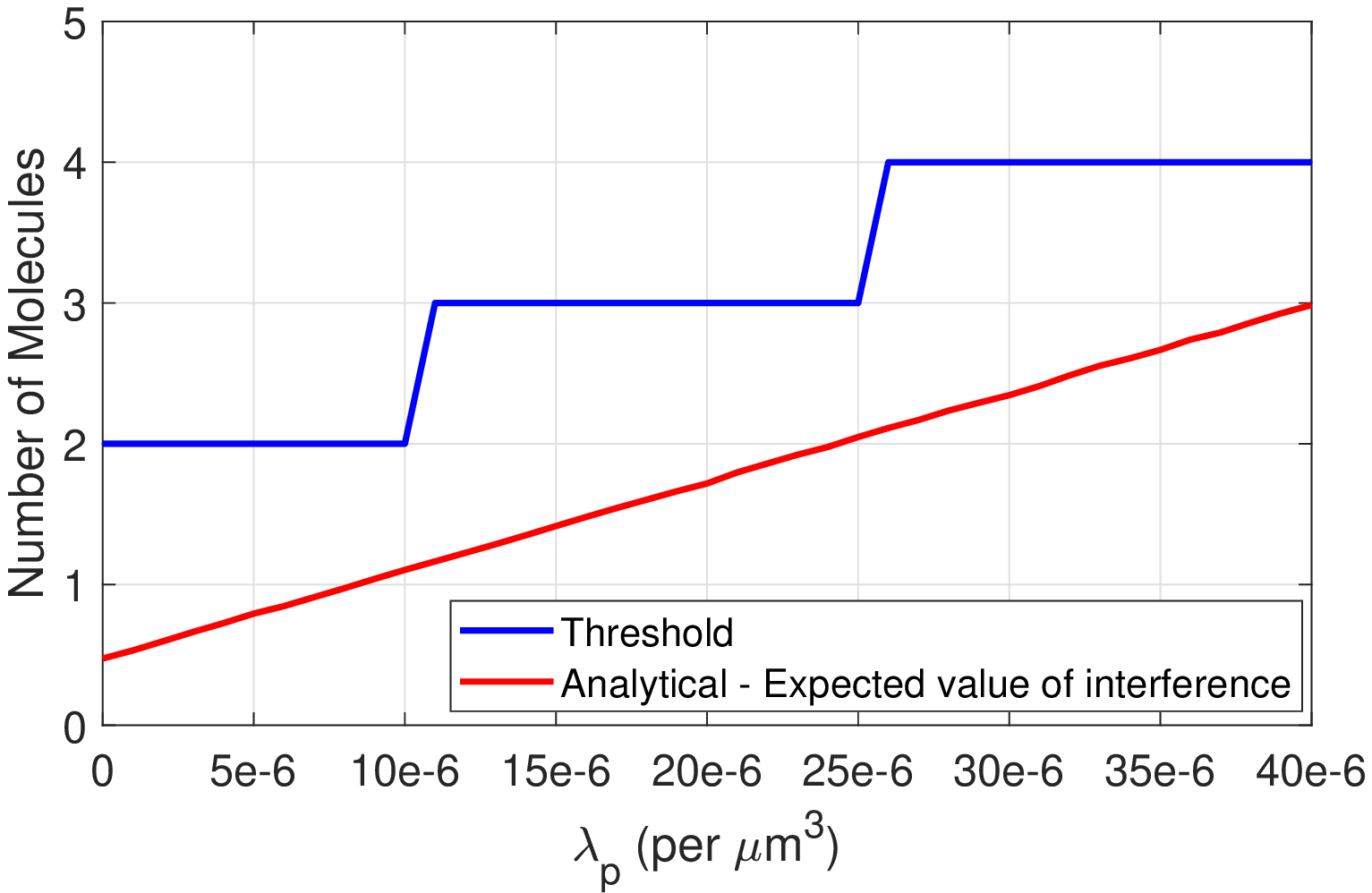} 
	\vspace{-3pt}
	\caption{Number of molecules as a function of $\lambda_\text{p}$.}
	\vspace{-8pt}
\end{figure}

\begin{figure}[tb!]
	\vspace{0pt}
	\centering
	\includegraphics[width =3.2in]{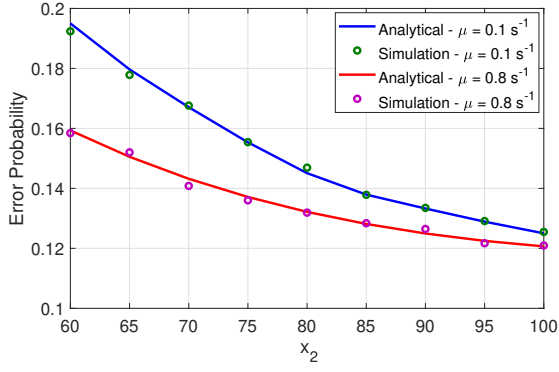} 
	\vspace{-3pt}
	\caption{Error probability as a function of $x_2$.}
	\vspace{-8pt}
\end{figure}

\begin{figure}[tb!]
	\vspace{0pt}
	\centering
	\includegraphics[width =3.2in]{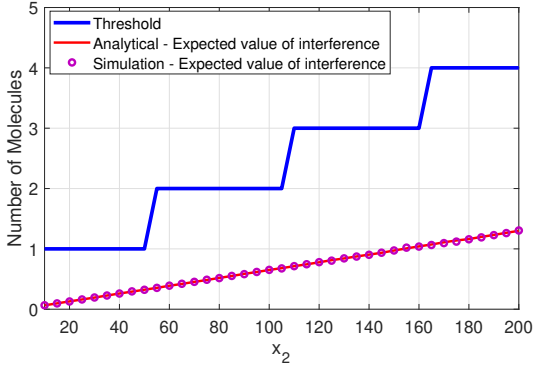} 
	\vspace{-3pt}
	\caption{Number of molecules as a function of $x_2$.}
	\vspace{-8pt}
\end{figure}

\section{Conclusions}
In this paper, we used tools from stochastic geometry to develop the first comprehensive framework for the modeling, analysis, and design of MC in bio-nanonetworks whose NMs form clusters around their respective FCs. In order to capture the coupling in the locations of NMs and their respective FCs, we modeled NMs as a PCP with the FCs forming the parent point process. This departs significantly from the known approaches that rely on PPP-based models and often ignore the complex dependence of detectors on interference. For the proposed model, we first characterized the distributions of the distances from a reference FC to various intra- and inter-cluster NMs. We also identified a specific structure for 3D TCPs that provided a remarkably simple expression for the distance distribution in the TPC setting. Then, based on the expected value of intra- and inter-cluster interferences, we proposed a simple detector for FCs that is suitable for biological applications. Using the developed tools, we also provided analytical expressions and easy-to-use approximations for the error probability of the proposed detector. Our analysis revealed that decreasing the intensity of cluster centers or the distance of the NM from the center of its cluster and also increasing the time slot duration or number of released molecules can improve the performance in terms of the error probability.

\appendices
\section{Proof of Theorem 1}
Defining $\mathbf{z} = \mathbf{x}+\mathbf{y} \in \mathbb{R}^3$, where $\mathbf{z} = (z_1,z_2,z_3)$ and $\mathbf{x} = (x_1,x_2,x_3)$, the conditional cumulative distribution function (CDF) of the distance $d = \|\mathbf{z}\|$ with realization $y=\sqrt{z_1^2+z_2^2+z_3^2}$ is
\begin{align}
\label{CDF}
&\mathbb{P}(d<y\mid \mathbf{x})= \int_{z_1=-y}^{y}\int_{z_2 = -\sqrt{y^2-z_1^2}}^{\sqrt{y^2-z_1^2}}\int_{z_3=-\sqrt{y^2-z_1^2-z_2^2}}^{\sqrt{y^2-z_1^2-z_2^2}}\nonumber\\&f_{\mathbf{Y}}(z_1-x_1,z_2-x_2,z_3-x_3)\mathrm{d}z_3\mathrm{d}z_2\mathrm{d}z_1\nonumber\\
&\stackrel{(a)}{=}\int_{z_1=-y}^{y}\int_{z_2 = -\sqrt{y^2-z_1^2}}^{\sqrt{y^2-z_1^2}}\int_{z_3=-\sqrt{y^2-z_1^2-z_2^2}}^{\sqrt{y^2-z_1^2-z_2^2}}\nonumber\\&f_{\mathbf{Y}}(z_1-\|\mathbf{x}\|,z_2,z_3)\mathrm{d}z_3\mathrm{d}z_2\mathrm{d}z_1=\mathbb{P}(d<y\mid \|\mathbf{x}\|) \nonumber\\&= F_d(y|{\|\mathbf{x}\|}),
\end{align}
where $(a)$ is due to the fact that $f_{\mathbf{Y}}(\mathbf{y})$ is rotationally invariant. It is notable that the result of \eqref{CDF} is dependent to the norm of $\mathbf{x}$. Then, by taking a derivative of the CDF $F_d(y|{\|\mathbf{x}\|})$, the conditional PDF $f_d(y|{\|\mathbf{x}\|})$ is obtained with the help of the Leibniz integral rule \cite{lebenan} and simplifications.

For the special case of TCP, by substituting the following $f_{\mathbf{Y}}$ from (1) as
\begin{align}
&f_{\mathbf{Y}}(z_1-\|\mathbf{x}\|,z_2,z_3) =\nonumber\\ &\frac{1}{(2\pi)^\frac{3}{2} \sigma^3}\exp\left(-\frac{(z_1-\|\mathbf{x}\|)^2+z_2^2+z_3^2}{2 \sigma^2}\right),
\end{align}
into \eqref{PCP_PDF}, we obtain
\begin{align}
&f_d(y|{\|\mathbf{x}\|}) =  \frac{2y}{(2\pi)^\frac{3}{2}\sigma^3}\exp\left(-\frac{y^2+\|\mathbf{x}\|^2}{2\sigma^2}\right) \int_{z_1=-y}^{y}\nonumber\\&\exp\left(\frac{\|\mathbf{x}\|z_1}{\sigma^2}\right)\int_{z_2 = -\sqrt{y^2-z_1^2}}^{\sqrt{y^2-z_1^2}}\frac{1}{\sqrt{y^2-z_1^2-z_2^2}}\mathrm{d}z_2\mathrm{d}z_1,
\end{align}
where $\int_{z_2 = -\sqrt{y^2-z_1^2}}^{\sqrt{y^2-z_1^2}}\frac{1}{\sqrt{y^2-z_1^2-z_2^2}}\mathrm{d}z_2 = \sin^{-1}\left(\frac{z_2}{\sqrt{y^2-z_1^2}}\right)$ $\bigg|_{-\sqrt{y^2-z_1^2}}^{\sqrt{y^2-z_1^2}} = \pi$. Then, we obtain
\begin{align}
f_d(y|{\|\mathbf{x}\|}) &= \frac{y}{\sqrt{2\pi}\sigma \|\mathbf{x}\|}\exp\left(-\frac{y^2+\|\mathbf{x}\|^2}{2\sigma^2}\right)\times\nonumber\\&\left[\exp\left(\frac{y\|\mathbf{x}\|}{\sigma^2}\right)-\exp\left(-\frac{y\|\mathbf{x}\|}{\sigma^2}\right)\right],
 \end{align}
which leads to the final result.
\section{Proof of Lemma 1}
The expected value of ${\cal I}_L^{\text{intra}}$ is obtained as
\begin{align}
\label{intra_cluster_omid}
&\mathbb{E}\left\{{\cal I}_L^{\text{intra}}\right\} = \mathbb{E}\left\{ \sum_{i=1}^{L-1}p_{iL}(\|\mathbf{y}_\mathbf{o}^i\|)X_{\mathbf{y}_\mathbf{o}^i}\right\}\nonumber\\&=\mathbb{E}\left\{X_{\mathbf{y}_\mathbf{o}^1}\right\}\mathbb{E}\left\{\sum_{i=1}^{L-1}p_{iL}(\|\mathbf{y}\|)\bigg|\|\mathbf{y}\|>r_0\right\}\nonumber\\ &=\left(\frac{1}{M}\sum_{j=1}^{M}{x_j}\right)\sum_{i=1}^{L-1}\frac{\int_{r_0}^{\infty}p_{iL}(y)f_{d}(y|0)\mathrm{d}y}{\int_{r_0}^{\infty}f_{d}(y|0)\mathrm{d}y} \nonumber\\&= \frac{\sum_{j=1}^{M}{x_j}}{M\int_{r_0}^{\infty}f_{d}(y|0)\mathrm{d}y}\int_{r_0}^{\infty}f_{d}(y|0)\sum_{i=1}^{L-1}p_{iL}(y)\mathrm{d}y,
\end{align}
where the term $\int_{r_0}^{\infty}f_{d}(y|0)\mathrm{d}y = \mathbb{P}(\|\mathbf{y}\|>r_0)$ is for the condition that NMs are outside the reference FC, i.e., the ball with radius $r_0$ centered at the origin.

Then, for $\mathbb{E}\left\{{\cal I}_L^{\text{inter}}\right\}$, we have
\begin{align}
\label{inter_cluster_omid}
&\mathbb{E}\left\{{\cal I}_L^{\text{inter}}\right\} = \mathbb{E}\left\{\sum_{\mathbf{x}\in\Phi_\text{p}}^{}\sum_{i=1}^{L}p_{iL}(\|\mathbf{x}+\mathbf{y}_\mathbf{x}^i\|)X_{\mathbf{x}+\mathbf{y}_\mathbf{x}^i}\right\}\nonumber\\&=\mathbb{E}\left\{X_{\mathbf{x}_1+\mathbf{y}_{\mathbf{x}_1}^1}\right\}\mathbb{E}\left\{\sum_{\mathbf{x}\in\Phi_{\rm p} }^{}\sum_{i=1}^{L}p_{iL}(\|\mathbf{x}+\mathbf{y}\|)\bigg|\|\mathbf{y}\|>r_0\right\}\nonumber
\end{align}
\begin{align}
&\stackrel{(a)}{=}\left(\frac{1}{M}\sum_{j=1}^{M}{x_j}\right)\mathbb{E}\Biggl\{\sum_{\mathbf{x}\in\Phi_{\rm p}}^{}\sum_{i=1}^{L}\biggl(\int_{r_0}^{\infty}p_{iL}(y)\times\nonumber\\&\frac{f_{d}(y|{\|\mathbf{x}\|})}{\int_{r0}^{\infty}f_d(y|{\|\mathbf{x}\|})\mathrm{d}y}\mathrm{d}y\biggr)\Biggr\} \stackrel{(b)}{=} 4\pi \lambda_\text{p} \left(\frac{1}{M}\sum_{j=1}^{M}{x_j}\right)\times\nonumber\\
&\int_{2r_0}^{\infty} \frac{1}{\int_{r0}^{\infty}f_d(y|x)\mathrm{d}y}\sum_{i=1}^{L}\left(\int_{r_0}^{\infty}p_{iL}(y)f_{d}(y|x)\mathrm{d}y\right) x^2\mathrm{d}x\nonumber\\&
= \frac{4\pi \lambda_\text{p}\sum_{j=1}^{M}{x_j}}{M} \int_{2r_0}^{\infty} \frac{x^2\int_{r_0}^{\infty} f_{d}(y|x)\sum_{i=1}^{L}p_{iL}(y)\mathrm{d}y}{\int_{r0}^{\infty}f_d(y|x)\mathrm{d}y} \mathrm{d}x,
\end{align}
where $(a)$ follows from the fact that the distance of each NM in the cluster $\mathbf{x} \in \Phi_{\text{p}}$ to the origin is i.i.d. with distribution $f_d(.|{\|\mathbf{x}\|})$ \cite{azimi_cluster1, afshang_cluster2} and the term $\int_{r_0}^{\infty}f_{d}(y|{\|\mathbf{x}\|})\mathrm{d}y = \mathbb{P}(\|\mathbf{y}+\mathbf{x}\|>r_0)$ is for the condition that NMs are outside the ball with radius $r_0$ centered at the origin. Then, $(b)$ follows from the Campbell's theorem for PPPs\cite{haenggi_book} and the fact that FCs have at least $2r_0$ distance from each other. 
\section{Proof of Lemma 2}
The LT of the interference is
\begin{eqnarray}
&{\cal L}_{{\cal I}_L} (s) = \mathbb{E}\left\{e^{-s{{\cal I}_L}}\right\} = \mathbb{E}\left\{e^{-s\left({{\cal I}_L^{\text{intra}}}+{{\cal I}_L^{\text{inter}}}\right)}\right\} \nonumber\\&= {\cal L}_{{\cal I}_L^{\text{intra}}} (s) {\cal L}_{{\cal I}_L^{\text{inter}}} (s),
\end{eqnarray}
where for ${\cal L}_{{\cal I}_L^{\text{intra}}}$, we have
\begin{align}
&{\cal L}_{{\cal I}_L^{\text{intra}}} (s) = \mathbb{E}\left\{e^{-s{{\cal I}_L^{\text{intra}}}}\right\}=\mathbb{E}\left\{e^{-s\sum_{i=1}^{L-1}p_{iL}(\|\mathbf{y}_\mathbf{o}^i\|)X_{\mathbf{y}_\mathbf{o}^i}}\right\} \nonumber\\&=\mathbb{E}\left\{ \mathop \prod \limits_{i=1}^{L-1} \frac{1}{M}\sum_{j=1}^{M}{e^{-s x_j p_{iL}{(\|\mathbf{y}\|)}}}\bigg| \|\mathbf{y}\|>r_0\right\}\nonumber\\&=\frac{1}{M^{L-1}}\left(\mathop \prod \limits_{i=1}^{L-1}\int_{r_0}^{\infty}\sum_{j=1}^{M} e^{-s x_j p_{iL}(y)} \frac{f_d(y|0)}{\int_{r_0}^{\infty}f_d(y|0)\mathrm{d}y}\mathrm{d}y\right)\nonumber\\&=\frac{1}{\left(M{\int_{r_0}^{\infty}f_d(y|0)\mathrm{d}y}\right)^{L-1}}\times\nonumber\\&\mathop \prod \limits_{i=1}^{L-1}\int_{r_0}^{\infty}\sum_{j=1}^{M} e^{-s x_j p_{iL}(y)} f_d(y|0)\mathrm{d}y,
\end{align}
where the condition that $\|\mathbf{y}\|$ is more than $r_0$ is applied. For ${\cal L}_{{\cal I}_L^{\text{inter}}}$, we have
\begin{align}
&{\cal L}_{{\cal I}_L^{\text{inter}}} (s) = \mathbb{E}\left\{e^{-s{{\cal I}_L^{\text{inter}}}}\right\}\nonumber\\&=\mathbb{E}\left\{e^{-s\sum_{\mathbf{x}\in\Phi_\text{p}}^{}\sum_{i=1}^{L}p_{iL}(\|\mathbf{x}+\mathbf{y}_\mathbf{x}^i\|)X_{\mathbf{x}+\mathbf{y}_\mathbf{x}^i}}\right\} \nonumber\\
&= \mathbb{E}\left\{ \mathop \prod \limits_{\mathbf{x}\in\Phi_{\rm p}}^{}\mathop \prod \limits_{i=1}^{L} \frac{1}{M}\sum_{j=1}^{M}{e^{-s x_j p_{iL}{(\|\mathbf{x}+\mathbf{y}\|)}}}\bigg| \|\mathbf{y}\|>r_0\right\}\stackrel{(a)}{=}\nonumber\\
& \mathbb{E}\left\{ \mathop \prod \limits_{\mathbf{x}\in\Phi_{\rm p}}^{}\frac{1}{M^L}\mathop \prod \limits_{i=1}^{L}\int_{r_0}^{\infty}\sum_{j=1}^{M}{e^{-s x_j p_{iL}{(y)}}}\frac{f_d(y|{\|\mathbf{x}\|})}{\int_{r_0}^{\infty}f_d(y|{\|\mathbf{x}\|})\mathrm{d}y}\mathrm{d}y\right\}\nonumber
\end{align}
\begin{align}
&\stackrel{(b)}{=}\exp \Biggl(-4\pi\lambda_\text{p} \int_{2r_0}^{\infty}\biggl(1-\frac{1}{\left(M{\int_{r_0}^{\infty}f_d(y|{\|\mathbf{x}\|})\mathrm{d}y}\right)^L}\times\nonumber\\
&\mathop \prod \limits_{i=1}^{L}\int_{r_0}^{\infty}\sum_{j=1}^{M}{e^{-s x_j p_{iL}{(y)}}}f_d(y|x)\mathrm{d}y\biggr)x^2\mathrm{d}x \Biggr),
\end{align}
where $(a)$ follows from the fact that the distance of each NM in the cluster with $\mathbf{x}\in \Phi_\text{p}$ to the origin is i.i.d. with distribution $f_d(.|{\|\mathbf{x}\|})$ \cite{azimi_cluster1, afshang_cluster2} and $(b)$ follows from the PGFL of PPPs \cite{haenggi_book} and the minimum distance between FCs, i.e., $2r_0$.

\section{Proof of Theorem 2}
From \eqref{BEP}, the conditional probability of error can be written as
\begin{align}
&\mathbb{P} ({\cal E}\mid X_{\mathbf{y}_0}=x_j)  =1- \mathbb{E}\Biggl\{\sum_{k=0}^{\text{th}_{j}-1}e^{-\left(p_{LL}(\|\mathbf{y}_0\|) x_j+{{\cal I}_L}+\lambda_0 T\right)}\times\nonumber\\
& \frac{(p_{LL}(\|\mathbf{y}_0\|) x_j+{{\cal I}_L}+\lambda_0 T)^k}{k!}\Biggr\}+\mathbb{E}\Biggl\{\sum_{k=0}^{\text{th}_{j-1}-1}\nonumber\\
& e^{-\left(p_{LL}(\|\mathbf{y}_0\|) x_j+{{\cal I}_L}+\lambda_0 T\right)}\frac{(p_{LL}(\|\mathbf{y}_0\|) x_j+{{\cal I}_L}+\lambda_0 T)^k}{k!}\Biggr\} \nonumber\\
&\stackrel{(a)}{=} 1-\mathbb{P}(H_j > p_{LL}(\|\mathbf{y}_0\|) x_j+{{\cal I}_L}+\lambda_0 T) \nonumber\\
&+ \mathbb{P}(H_{j-1} > p_{LL}(\|\mathbf{y}_0\|) x_j+{{\cal I}_L}+\lambda_0 T) \nonumber\\
&\stackrel{(b)}{\leq} 1 - \biggl(1 - \mathbb{E}\left\{\left(1-e^{-{\eta}_j (p_{LL}(\|\mathbf{y}_0\|) x_j+{{\cal I}_L}+\lambda_0 T)}\right)^{\text{th}_j}\right\}\biggr) \nonumber\\&+ \biggl(1 - \mathbb{E}\left\{\left(1-e^{-{\eta}_{j-1} (p_{LL}(\|\mathbf{y}_0\|) x_j+{{\cal I}_L}+\lambda_0 T)}\right)^{\text{th}_{j-1}}\right\}\biggr)\nonumber\\
&\stackrel{(c)}{=} 1- \sum_{k=1}^{{\text{th}}_j} (-1)^{k+1} {{{\text{th}}_j} \choose {k}} e^{-{\eta}_j (p_{LL}(\|\mathbf{y}_0\|) x_j+\lambda_0 T) k}{\cal L}_{{\cal I}_L}({\eta}_jk) \nonumber\\&+ \sum_{k=1}^{{\text{th}}_{j-1}} (-1)^{k+1} {{{\text{th}}_{j-1}} \choose {k}} e^{-{\eta}_{j-1} (p_{LL}(\|\mathbf{y}_0\|) x_j+\lambda_0 T) k}\nonumber\\&\times{\cal L}_{{\cal I}_L}({\eta}_{j-1}k),
\end{align}
where $(a)$ comes from chi-squared distributions $H_{j-1}$ and $H_{j}$ with $2 \text{th}_{j-1}$ and $2 \text{th}_{j}$ degrees of freedoms, respectively, and $(b)$ is from the Alzer's lemma \cite{alzer} on chi-squared distributions $H_{j-1}$ and $H_{j}$ with parameters ${\eta}_{j-1} = {\text{th}_{j-1}!}^{-\frac{1}{\text{th}_{j-1}}}$ and ${\eta}_j = {\text{th}_j!}^{-\frac{1}{\text{th}_j}}$, respectively. Also, $(c)$ is from the binomial expansion and the definition of the LT. 

\section{Proof of Lemma 3}
Under this case, we have $\text{th}_1= 1$, which leads to the final error probability result from \eqref{error}-\eqref{BEP}. In the following, we prove that $\text{th}_1 = 1$ iff \eqref{xi} holds. From \eqref{threshold}, $\text{th} < 1$ if
\begin{align}
\frac{p_{LL}(\|\mathbf{y}_0\|) \xi }{{\ln}\left(p_{LL}(\|\mathbf{y}_0\|) \xi+\mathbb{E}\left\{{{\cal I}_L}\right\}+\lambda_0 T\right)-{\ln}\left(\mathbb{E}\left\{{{\cal I}_L}\right\}+\lambda_0 T\right)} < 1,
\end{align}
which leads to
\begin{align}
\frac{p_{LL}(\|\mathbf{y}_0\|) \xi+\mathbb{E}\left\{{{\cal I}_L}\right\}+\lambda_0 T}{\mathbb{E}\left\{{{\cal I}_L}\right\}+\lambda_0 T} > e^{{p_{LL}(\|\mathbf{y}_0\|) \xi}},
\end{align}
which can be simplified to $\xi < \xi_0$, where $\xi_0$ is the solution of
\begin{align}
\label{phase1}
\frac{p_{LL}(\|\mathbf{y}_0\|) \xi_0+\mathbb{E}\left\{{{\cal I}_L}\right\}+\lambda_0 T}{\mathbb{E}\left\{{{\cal I}_L}\right\}+\lambda_0 T} = e^{{p_{LL}(\|\mathbf{y}_0\|) \xi_0}}.
\end{align}
Then, \eqref{phase1} can be rewritten as
\begin{align}
\label{phase2}
&-({p_{LL}(\|\mathbf{y}_0\|) \xi_0+\mathbb{E}\left\{{{\cal I}_L}\right\}+\lambda_0 T}) e^{-({p_{LL}(\|\mathbf{y}_0\|) \xi_0+\mathbb{E}\left\{{{\cal I}_L}\right\}+\lambda_0 T})} \nonumber\\&= -({\mathbb{E}\left\{{{\cal I}_L}\right\}+\lambda_0 T}) e^{-({\mathbb{E}\left\{{{\cal I}_L}\right\}+\lambda_0 T})}.
\end{align}
From the Lambert function definition \cite{lambert}, \eqref{phase2} leads to
\begin{align}
&-({p_{LL}(\|\mathbf{y}_0\|) \xi_0+\mathbb{E}\left\{{{\cal I}_L}\right\}+\lambda_0 T}) = \nonumber\\&{\cal W}\left(-({\mathbb{E}\left\{{{\cal I}_L}\right\}+\lambda_0 T}) e^{-({\mathbb{E}\left\{{{\cal I}_L}\right\}+\lambda_0 T})}\right),
\end{align}
and then
\begin{align}
&\xi_0 = - \frac{\mathbb{E}\left\{{\cal I}_{L}\right\}+\lambda_0 T}{p_{LL}(\|\mathbf{y}_0\|)} -\frac{1}{p_{LL}(\|\mathbf{y}_0\|)} \times\nonumber\\&{\cal W}\left(-({\mathbb{E}\left\{{\cal I}_{L}\right\}+\lambda_0 T})e^{-({\mathbb{E}\left\{{\cal I}_{L}\right\}+\lambda_0 T})}\right),
\end{align}
which completes the proof.

\end{document}